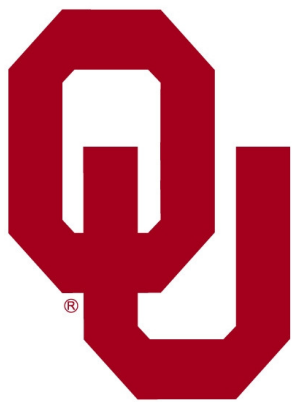

UNIVERSITY OF OKLAHOMA

TECHNICAL REPORT

---

# Performance Analysis of Message Dissemination Techniques in VANET using Fog Computing

___________________________________________________________

## Anirudh Paranjothi

anirudh.paranjothi@ou.edu

**BASED ON THE PAPERS:**

[1] L. Gao, T. H. Luan, S. Yu, W. Zhou and B. Liu, "FogRoute: DTN-Based Data Dissemination Model in Fog Computing," in *IEEE Internet of Things Journal*, vol. 4, no. 1, pp. 225-235, 2017. doi: 10.1109/JIOT.2016.2645559

[2] Y. Bi. "Neighboring vehicle-assisted fast handoff for vehicular fog communications." in *Peer-to-Peer Networking and Applications*, vol. 11, pp. 1-11, 2018. doi: 10.1007/s12083-017-0570-8

**10/05/2018**

# Abstract

Vehicular Ad-hoc Networks (VANET) is a derived subclass of Mobile Ad-hoc Networks (MANET) with vehicles as mobile nodes. VANET facilitate vehicles to share safety and non-safety information through messages. Safety information includes road accidents, natural hazards, roadblocks, etc. Non-safety information includes tolling information, traveler information, etc. The main goal behind sharing this information is to enhance road safety and reduce road accidents by alerting the driver about the unexpected hazards. However, routing of messages in VANET is challenging due to packet delays arising from high mobility of vehicles, frequently changing topology and high density of vehicles, leading to frequent route breakages and packet losses. This report summarizes the performance analysis of safety and non-safety message dissemination techniques in VANET based on the fog computing technique. Three main metrics to improve the performance of message dissemination are: 1) delay, 2) probability of message delivery, and 3) throughput. Analysis of such metrics plays an important role to improve the performance of existing message dissemination techniques. Simulations are usually conducted based on the metrics using ns-2 and Java discrete event simulator. The above three performance metrics and results published in literature help one to understand and increase the performance of various message dissemination techniques in a VANET environment.

# TABLE OF CONTENTS

# 1. INTRODUCTION

Vehicular ad-hoc networks (VANET) has evolved from mobile ad hoc networks (MANET) with distinguished characteristics like high mobility and rapid change in topology. VANET allow the vehicles to communicate with each other and exchange safety as well as non-safety information between the vehicles as messages [1]. Safety information includes road accident, roadblock, accident information, etc. Non-safety information includes tolling information, entertainment, etc. A report given by the association for safe and international road travel (ASIRT) concluded that nearly 1.25 million people die in road crashes each year, and distracted driving is one of the major reasons for road crashes [2]. As a result, VANET emerged as the promising solution with a motivation to improve the road safety by reducing road accidents.

Vehicle to vehicle communication (V2V) and vehicle to infrastructure communication (V2I) are the two communication techniques used in VANET. V2V allow the vehicles to communicate with each other directly using a multi-hop technique as long as the vehicles are in the transmission range of each other. An advantage of V2V communication is reduced communication overhead. However, it is not suitable for the long distance communication. V2I communication allows the vehicles to communicate with each other over a long distance using a multi-hop technique with the help of roadside infrastructure like road side units (RSU), etc. [3]. An advantage of V2I communication is providing support for the long distance communications. However, a considerable amount of communication overhead is involved in the transmission of messages. V2V and V2I communications are also known as short distance and long distance communications respectively. A sample scenario of V2V and V2I communication is depicted in Fig. 1.

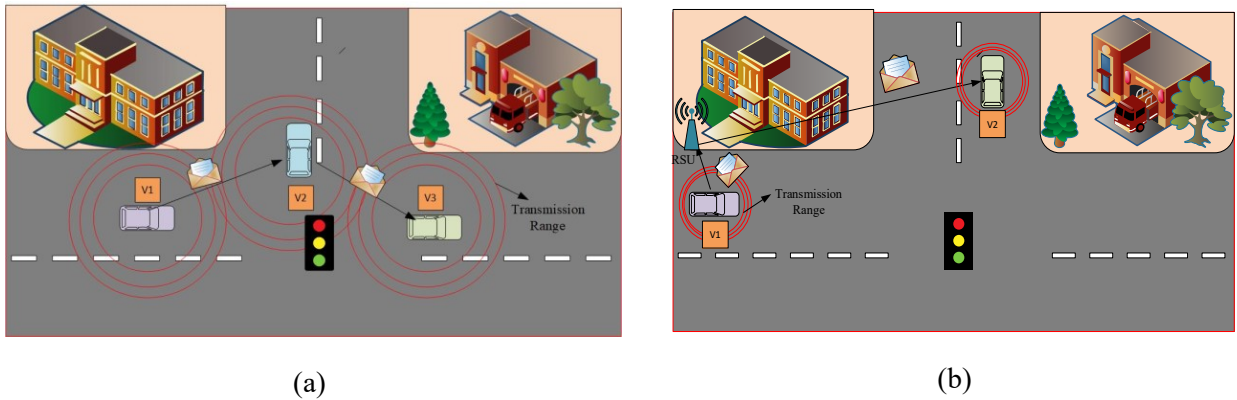

(a)    (b)

Fig. 1: Two types of communication in VANET: a) V2V communication, b) V2I communication.

V2V and V2I communications in VANET depends on the dedicated short range communication (DSRC) protocol. DSRC consists of a set of protocols for transmitting safety and non-safety information between vehicles and between vehicles and RSU. The federal communication commission (FCC) set aside 75 MHz bandwidth of 5.9 GHz (5.850 GHz to 5.925 GHz) band for vehicular communication [4], represented in Fig. 2. DSRC has one control channel and six service channels for communication, in which the control channel is used to transmit safety information such as road accidents, natural hazards, etc. and the service channels are used to transmit non-safety information such as parking information, personal messages, etc. However, the performance of DSRC significantly decreases as the number of vehicles increases in the system. For example, regions like Manhattan are always congested with more number of vehicles at most all times resulting in the increase of load on DSRC spectrum, leading to instability in DSRC.

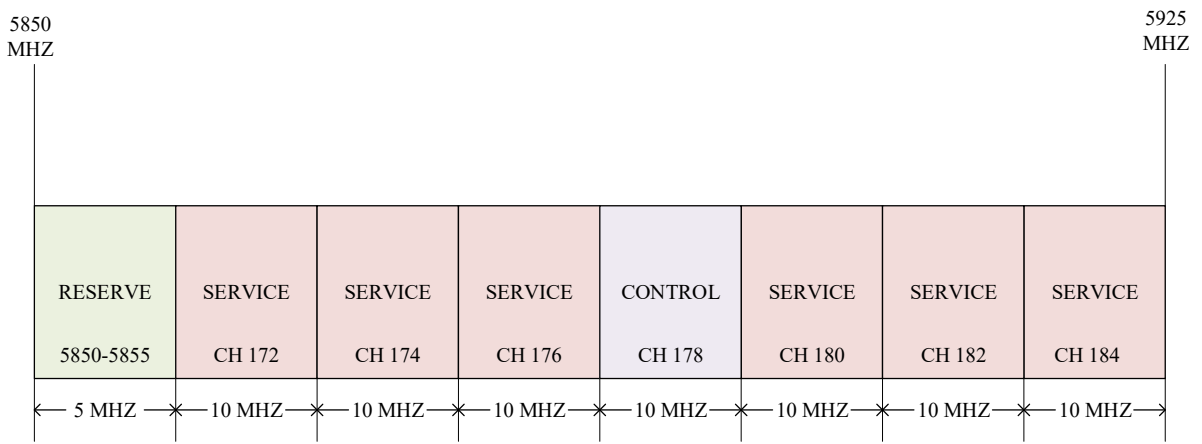

Fig. 2: DSRC spectrum with one control channel and six service channels.

In a VANET environment, information is transmitted among the vehicles in terms of messages. However, the current V2V and V2I communications do not guarantee the message delivery due to the instability of DSRC, resulting in messages being dropped before reaching the destination. Thus, the emergence of a new paradigm (Fog computing) is essential to guarantee the message delivery.

Fog computing (also known as edge computing) considered a new revolutionary way of thinking in wireless networking [5]. It is an extension of cloud computing where computations are performed at the edge of the network. Any real-world objects which can acquire the properties storage, computing, and network connectivity can be formed as a fog node for a time period ($t$), resulting in rapid dissemination of messages between the vehicles [6]. In addition, fog computing also offers special services including location awareness, ultra-low frequency, and context information.

Being able to extend the cloud service to an edge of the network is the remarkable characteristic of fog computing. By pooling the local services, fog computing enables control, computation, storage, and communication at the proximity of end users. By adding a resource rich layer between cloud and end devices, fog computing meets the challenges in high performance, interoperability, low latency, high reliability, mobility, and high security.

In fog computing, the extent of network transmission and the time required for data transfer are reduced as the network edge devices consume the data. The fog model can ease network bandwidth bottlenecks and adequately meet the needs of latency sensitive applications. It consists of many fog nodes, which includes: 1) virtualized edge data centers, 2) network edge devices, and 3) management systems. Fog nodes connect with users and end devices by wireless connections like Wi-Fi, 4G, Bluetooth, etc. in order to provide storage, computation, and computing services.

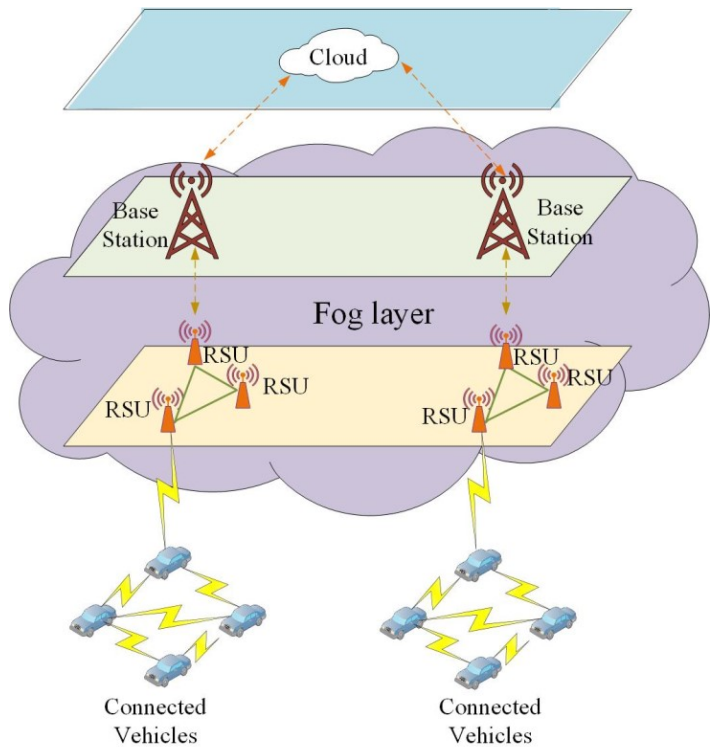

Fig. 3: General architecture of vehicular fog computing (VFC).

Fog Computing in a vehicular environment is commonly termed as vehicular fog computing (VFC), represented in Fig. 3. In VFC, vehicles are considered as infrastructure to make up most utilization of computational resources and vehicular communications. The main objective of VFC is to utilize a large number of near-user edge devices or end-user clients to carry computation and communication. Besides the cloud characteristics, like providing application, storage and computing services to end users, VFC differentiates itself from existing models with its dense geographical distribution and proximity to end users. Thus, VFC provides low-latency at most all times in vehicular communications compared to existing techniques.

The *objectives* of this report are:

- Analyzing the performance of message dissemination techniques in fog computing discussed by Gao *et al.* [7] and Bi [8] based on the metrics such as delay, probability of message delivery, and throughput.
- Reviewing the simulation results discussed by Gao *et al.* [7] and Bi [8].
- Providing the findings and future directions of methods discussed by Gao *et al.* [7] and Bi [8].

Rest of the report is organized as follows: First, the detail information about the proposed methods are discussed in Section 2. Then, analysis of method 1, FogRoute and method II, cross-layer and neighboring vehicle fast handoff (CVFH) are carried out in Sections 3 and 4, respectively. After the analysis part, simulation results are given in Section 3.3 and 4.3. The findings of FogRoute and CVFH are discussed in Section 5. Finally, the report concludes with future research directions in Section 6.

## 2. DESCRIPTION OF METHODS

Two different message dissemination techniques using fog computing (i.e., FogRoute and CVFH) have been studied and described in subsections 2.1 and 2.2, respectively.

### 2.1 FogRoute

In FogRoute, a hybrid data dissemination model using fog computing is proposed, which takes advantage of delay tolerant networking (DTN) and cloud techniques in the data dissemination process [7]. The hybrid data dissemination process of FogRoute is represented in Fig. 4. It consists of a cloud server, fog servers, and mobile devices such as vehicles, users, etc. Cloud server acts as a control plane to compare the contents of fog servers and determine the fog server needs to be updated with the required content. Fog servers act as the data plane to disseminate the content to the mobile devices such as vehicles, users, etc.

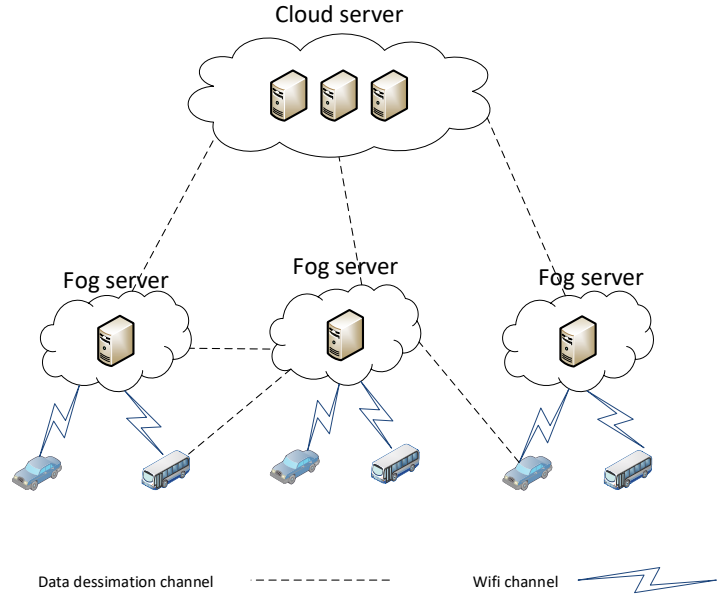

Fig. 4: Proposed architecture of FogRoute, a hybrid data dissemination model using fog computing.

FogRoute consists of the following three tables to store information about the list of mobile devices, fog servers, and its contents [7]:

- Global Content List Table (GC): GC is used to store the list of all public contents in fog servers, which includes fog servers id, content id, date of update, and validation time.
- Fog Server List Table (FC): FC is used to store the list of all available fog servers, which includes fog servers id, content id, and mobile devices id.
- Mobile Device Movement Pattern Table (MDMV): MDMV is used to store the list of all mobile devices id, fog server id, linked time, geographic movement pattern, and social attribute.

Following messages are disseminated between the cloud provider, fog servers and mobile devices in the data dissemination process of FogRoute [7]:

- Hello message: Hello is an update message exchanged between a fog server and its cloud provider, which includes content id and associated mobile device id.
- Data Dissemination Request Message: hen a cloud provider finds the content to be updated in the target fog server ($FS_t$), a data dissemination request message is sent from the cloud provider to the remaining fog servers (i.e., excluding the target fog server) to disseminate the content through the mobile devices attached to it.
- Data Dissemination Accept Message: When the fog server completes the data dissemination process to the mobile device attached to it, data dissemination accept message is sent from the fog server to its cloud provider.

- Data Dissemination Decline Message: When the fog server does not complete the data dissemination process to the mobile device attached to it, data dissemination decline message is sent from the fog server to its cloud provider.
- Data Dissemination Acknowledgement Message: When the target fog server ($FS_t$) receives the required content from the mobile device, a data dissemination acknowledgment is sent from the target fog server to its cloud provider.

Data flow and mobile device selection (if more than one mobile device is attached to the fog server) processes are briefly explained in Sec. 3.1

*2.2 CVFH*

CVFH is proposed to disseminate messages between the vehicles to perform a successful handoff from one access point (AP) to another. Three vehicles and two APs are used in CVFH [8], represented in Fig. 5:

- Current Vehicle (CV): CV is the vehicle that is going to experience the handoff.
- Neighbor Vehicles (NV): NVs are the vehicles in the communication range of CV.
- Neighbor Assisted Vehicle (NAV): Among the available NVs, CV selects the most qualified node to perform the handoff. The selected neighbor (NV) is known as NAV.
- Serving AP (SAP): SAP is the AP to which CV is currently connected.
- Target AP (TAP): TAP is the AP to which CV is going to connect after a successful handoff process.

In CVFH, CV communicates with SAP to obtain the received signal strength indication (RSSI) and packet loss rate (PLR). When the RSSI decreases and PLR increases continuously for a time period ($t$), CV prepares for the handoff process and disseminates neighbor request message to all the neighbors. Upon receiving the request message, only the qualified neighbors can reply it. The qualified neighbors should satisfy the following conditions [8]:

- The SAP of the neighbor should be different from the SAP of CV in one hop situation such that CV and NAV are in the communication range of each other.
- The position of a neighbor should be ahead of CV, which ensures CV is moving closer towards TAP.
- Only the neighbor who has not received any neighbor reply message can communicate with CV.

Once CV receives the neighbor reply message from the qualified neighbor, it validates the reply message to find if the neighbor is an AP or a vehicle and performs handoff once CV goes out of the range of SAP, briefly explained in Sec. 4.1.

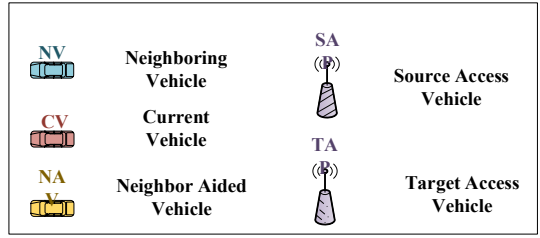

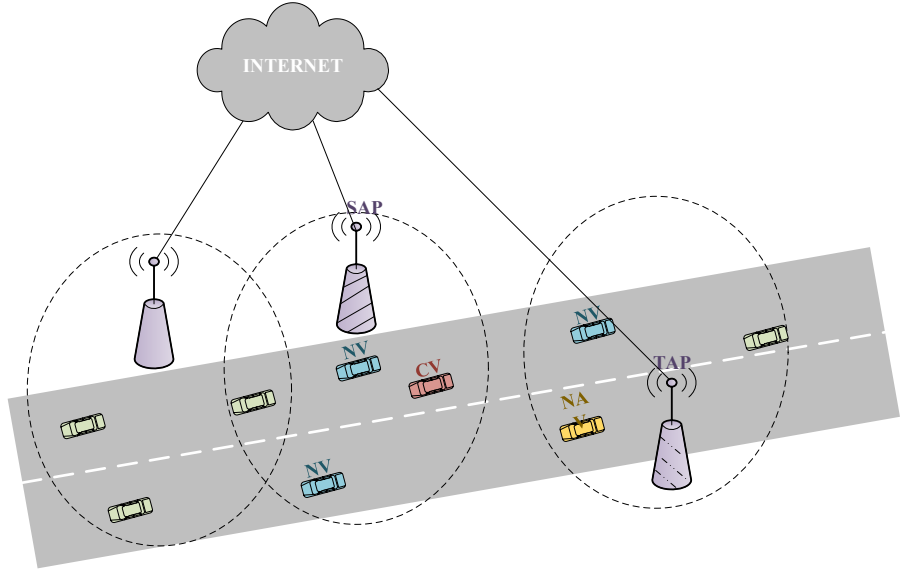

Fig. 5: Proposed architecture of a VFC based model CVFH.

## 3. ANALYSIS OF FOGROUTE

This section presents the analysis of FogRoute based on the following subsections: 1) algorithms (subsection 3.1), describes the set of instructions involved in the data dissemination process, 2) performance evaluation (subsection 3.2), illustrates the metrics considered to evaluate data dissemination protocol, and 3) simulation results (subsection 3.3), describes the results various simulations performed based on ns-2 simulator.

**Notations used in Algorithm 1:**

*CP*   = Cloud Provider

*GC*   = Global Content List Table

*FS*   = Fog Server List Table

$FS_i$   = Fog Server $i$

$FS_t$   = Target Fog Server $t$

$CR_n$   = Carrier or Mobile Device used to deliver the required content (C)

$C$      = Content to be delivered to the target fog server

## 3.1 Algorithms

| Algorithm 1: Data flow control algorithm |
| --- |

1.   *CP* monitors *GC* and *FS*
2.   **if** *CP* finds *C* in $FS_t$ has to be updated **then**
3.       **if** no other $FS_i$ has the content **then**
4.           *CP* sends *C* to $FS_t$
5.           *CP* update *GC* and *FS* table
6.       **else**
7.           **if** list of $FS_i$ has the *C* to update $FS_t$ **then**
8.               **if** $CR_n$ associated with $FS_i \geq 0$ **then**
9.                   Select suitable $CR_n$ [ described in Algorithm 2]
10.                  *CP* sends *data_dis_req* message to $FS_i$
11.                  Go to step 19
12.               **else**
13.                  *CP* sends *C* to $FS_t$
14.           **else**
15.               Go to 1
16.           **endif**
17.           **endif**
18.       **endif**
19.       **if** ($FS_i$ receives *Data_dis_req* message from *CP*) **then**
20.           Send *C* to $FS_i$ , $FS_t$ to $CR_n$
21.           **if** ($FS_i$ transfers *C* to $CR_n$) **then**
22.               $FS_i$ sends *Data_dis_accep* message to *CP*
23.           **else**
24.               $FS_i$ sends *Data_dis_decline* message to *CP*
25.       **else**
26.           Wait for *Data_dis_req* message from *CP*
27.       **endif**
28.       **endif**
29.   **if** $FS_t$ receives *C* from $FS_i$ **then**

| | |
|---|---|
| 30. | $FS_t$ send *Data_dissem_ack* message to *CP* |
| 31. | **else** |
| 32. | Wait for *Data_dissem_ack* message from $FS_t$ |
| 33. | **endif** |
| 34. | **if** *CP* receives *Data_dissem_ack* then |
| 35. | *CP* update *GC* and *FS* table |
| 36. | **else** |
| 37. | **if** *CP* receives *Data_dissem_decline* message then |
| 38. | Go to step 4 |
| 39. | **else** |
| 40. | *CP* waits for the message from $FS_t$ |
| 41. | **endif** |
| 42. | **endif** |
| 43. | **else** |
| 44. | Repeat from step 1 |
| 45. | **endif** |

Algorithm 1 describes the flow of the data dissemination process in FogRoute. When the cloud provider finds the content to be updated in the target fog server based on the comparison of global content list table and fog server list table, cloud provider checks the remaining fog servers for the required content. If none of the fog servers has the required content, cloud provider directly disseminates the content to the target fog server and updates the global content list table. But, if the cloud provider finds that the remaining fog servers have the content to be updated, it checks for the number of mobile devices ($n$) such as vehicles in the range of fog server to update the content. This leads to two possible cases: 1) one or more mobile devices in the range of fog server ($n>0$) and 2) no mobile devices in the range of fog server ($n=0$).

After finding the suitable fog servers to disseminate the content, cloud provider sends a data dissemination request message to the selected fog servers. Upon receiving a request message, fog server disseminates the content to the mobile devices attached to it and sends a data dissemination to accept message to the cloud provider. After receiving the content from the mobile device, the target fog server sends a data dissemination acknowledgment message to the cloud provider. Failure to disseminate the content results in sending a data dissemination decline message to the cloud provider.

**Notations used in Algorithm 2:**

| | |
|---|---|
| $T_d$ | = Affordable delay time of a content ( c ) |
| *MDMP* table | = Mobile Device Movement Pattern table |
| *Mob_list$_t$ (i)* | = list of all mobile devices attached to targeted fog server FS(t), where i is the id of each mobile devices |
| *new_mob_list$_t$(i)* | = new list of all mobile devices statisfies the condition [∵ *mob_conntime$_t$* (i) > *avg_time$_t$(i)*] |
| *mob_conntime$_t$(i)* | = connection time of mobile devices attached to the target fog server $FS_t$ [∵ $mob\_conntimet(i) = \frac{size_c}{speed_i}$ ] |
| *deli_time$_i$* | = average content delivery time of mobile devices (i) to the target fog server ($FS_t$) |
| *deli_prob* | = probability of content delivery |
| *avg_conntime$_t$(i)* | = average conectionn time between the mobile device (i) and target fog server ($FS_t$) |

---

Algorithm 2: Carrier selection algorithm

---

1. *CP* determines $T_d$ of *C* to be transmitted
2. *CP* checks *FS* tables and *MDMP* table to find *mob_list(i)*
3. *CP* checks the *avg_conntime$_t$(i)*
4. **for** all mobile devices (i.e. *md$_i$*) ∈ *mob_list$_t$(i)* do
5.    **if** (*mob_conntime$_t$(i)* > *avg_conntime$_t$(i)*) then
6.       *New_mob_list$_t$(i)* = *mobiledevice(i)* (i.e., *md$_i$*)
7.    **endif**
8. **endfor**
9. **for** all *md$_{si}$* ∈ *new_mob_list$_t$(i)* do
10.    Calculate *deli_time$_i$*
11.    **if** (*deli_time$_t$(i)* ≤ $T_{delay}$ ) then
12.       $CR_n$ = *md$_{si}$* (or) $CR_{si}$    [ where $CR_n$ = $CR_{s1}$, $CR_{s2}$, …… $CR_{sx}$ ; where x is the number of scheduled mobile devices selected for service]
13.    **endif**
14. **endfor**

15. **for** all $md_{nsi} \in new\_mob\_list_t(i)$ do
16.     Reorder mobile devices based on *deli_prob* [described in Section 3.2]
17. **endfor**
18. *CP* selects top y mobile devices based on *deli_prob* for content dissem
19. *CP* selects number of y based on $\frac{\sum_{i=1}^{x} deli\_time_{si} + \sum_{i=1}^{n} deli\_time_{nsi}}{x+y} < T_{delay}$
20. Update $CRn = x + y$

Algorithm 2 describes the carrier selection process of FogRoute protocol. Cloud provider determines the affordable time delay for the required content, checks for the list of mobile devices connected to the target fog server and calculates average connection of mobile devices using FS table and MDMP table. The mobile devices with shorter connection time are filtered out as they cannot upload the required content to the target fog server and the new list is created with mobile devices having high average connection time.

The new mobile devices list is further classified into two categories: 1) scheduled visits and 2) non-scheduled visits. Scheduled visits include the list of mobile devices having a predetermined visit to the fog server such as buses, airport shuttles, etc. Upon creation of scheduled visit list, the list of mobile devices having delivery time less than the affordable delay time are filtered out and added into the data dissemination carrier list. The delivery time is calculated based on the speed and direction of the mobile device (Sec. 3.2). Non-scheduled visits include the list of mobile devices not having a predetermined visit to the fog server. Upon creation of the non-scheduled list, the list of mobile devices having high delivery probability are filtered out and added to the data dissemination carrier list. The selection of mobile devices from the non-scheduled lists are briefly discussed in Sec. 3.2.

*3.2 Performance Evaluation*

Performance of scheduled mobile devices (i.e., devices having predefined scheduled time) is calculated based on the delivery time. The selected mobile devices are added to the data dissemination carrier list (carrier$_{s1}$, carrier$_{s2}$,…, carrier$_{sx}$), where $x$ is the number of mobile devices added to the carrier list. Performance of non-scheduled mobile devices (i.e., devices not having predefined scheduled time) is calculated based on the delivery probability. The selected mobile devices with high delivery probability are added to the data dissemination carrier list.

For each mobile device, cloud provider collects the contact frequency with a target fog server, geographic location, and three last visits. Based on the three last visits to the target server, average speed and direction of each mobile devices are calculated, which in turn are used to calculate the delivery time of the mobile device (i.e., DeliTime$_m$ = Speed$_m$/ Direction$_m$). The devices having delivery time greater than the affordable delay time are filtered out from the non-schedule mobile list, and for the remaining mobile devices the delivery probability is calculated as [7]:

$$DeliProb_m = \frac{ConFre_m}{\sum_{i=1}^{n} ConFre_i} \times \left(1 - \frac{DeliTime_m}{\sum_{i=1}^{n} DeliTime_i}\right) \qquad (1)$$

Where, $DeliProb_m$ = delivery probability of a mobile device, $ConFre_m$ = contact frequency of a mobile device, $\sum_{i=1}^{n} ConFre_i$ = sum of contact frequency of all mobile devices, $DeliTime_m$ = delivery time of a mobile device, and $\sum_{i=1}^{n} DeliTime_i$ = sum of delivery time of all mobile devices.

Based on the delivery probabilities, cloud provider selects top y devices, where y is the number of devices satisfies the following condition [7]:

$$\frac{\sum_i^x DeliTime_{Si} + \sum_{i=1}^{y} DeliTime_{NSi}}{x+y} < T_{delay} \qquad (2)$$

Where, $DeliTime_{Si}$ = delivery time of a scheduled mobile and $DeliTime_{NSi}$ = delivery time of a non-scheduled mobile device.

From Eqns. (1) and (2), $x + y$ mobile devices are selected as carriers to disseminate the data.

*3.3 Simulation Results*

DTN based data dissemination is simulated using a data set with 370 taxi cabs movement record for 6 months in the city of Rome, represented in Fig. 6. For every 7 seconds, the location of the taxi is updated to a server through an Android OS tablet running on an app equipped in each taxi. The data collected from February 1st, 2014 to March 2nd, 2014 is used for the simulation. Apart from taxis, the bus traces in the city of Rome are also collected. In the simulation areas, 50 fog servers are deployed. A Java based event driven simulator is implemented to simulate taxis and busses movements.

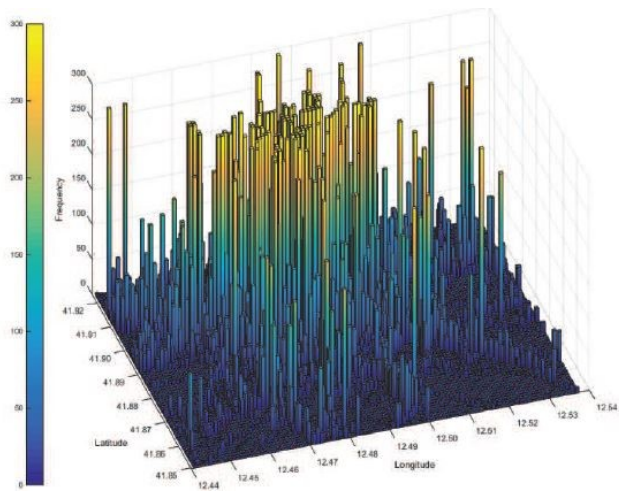

Fig. 6: Frequency of 370 taxi cabs reported locations in Rome.

The performance of the FogRoute is evaluated through the following tests: 1) testing the changes in expected delay to the delivery success ratio, and 2) testing the convergence time of data dissemination for each fog server.

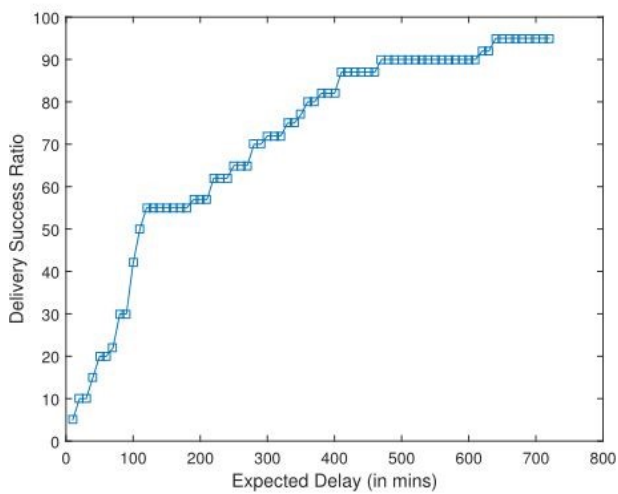

Fig. 7: Delivery ratio of FogRoute for various expected delays.

Delivery success ratio is one of the main factors to be considered while designing a data dissemination model. For this experiment, Fog server and the content to be disseminated are chosen randomly. The obtained results show that the changes in expected delay effects the delivery success ratio to a greater extent, represented in Fig. 7. The obtained average delivery success ratio is 50%, 80%, 90% and 95% when the expected delay is increased to 2 hrs, 6 hrs, 8 hrs and 10 hrs respectively. These results give an idea of how to use the proposed model. For instance, if more delay in content dissemination is acceptable, then the DTN based data dissemination is an excellent choice. If the delay in content dissemination is not acceptable, then it is sufficient to use DTN based data dissemination, but a backup of Cloud based data transmission channel has to be maintained.

The time spent on each Fog server to receive all contents is the convergence time. For this experiment, destinations to Fog servers are set by collecting 20 contents from the list. The total time taken by Fog servers to receive these 20 contents is the convergence time. The time taken for 7, 15 and 27 Fog servers to complete data dissemination is 10 min, 1 hr and 2 hrs time respectively, represented in Fig. 8. Altogether, 46 Fog servers were able to reach convergence in 11hrs, whereas 4 Fog servers were not able to converge in 12 hrs. From the above results, it can be seen that within a day, 92% of contents could be disseminated to every Fog server. But only 4% of Fog servers were not able to successfully update their content list, resulting in a requirement to have a cloud based data dissemination deployment.

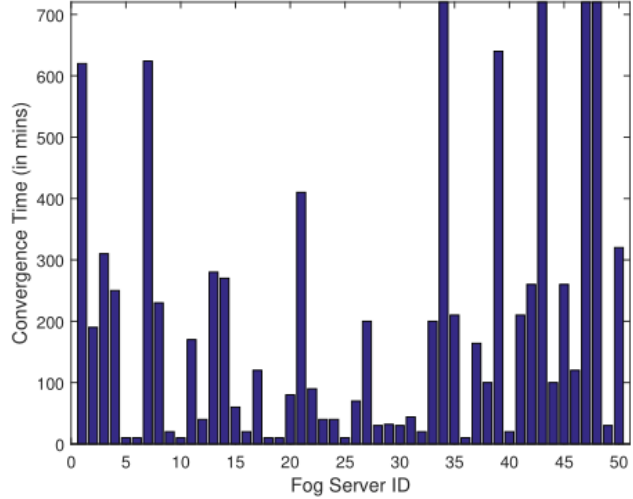

Fig. 8: Convergence time of each fog servers of FogRoute.

## 4. ANALYSIS OF CVFH

This section presents the analysis of proposed CVFH based on the following subsections: 1) algorithms (Subsection 4.1), describes the set of instructions involved in the handoff process, 2) performance evaluation (Subsection 4.2), illustrates the metrics considered to evaluate CVFH protocol , and 3) simulation results (Subsection 4.3), describes the results various simulations performed based on Java discrete event simulator.

**Notations used in Algorithm 3**

$CV$     = Current vehicle

$NV$     = Neighboring vehicle

$NAV$     = Neighbor assisted vehicle

$RSSI$     = Received Signal Strength Indication

$PLR_{CV}$ = Packet loss rate of $CV$

$PLR_{TH}$ = Predefined threshold (i.e., Acceptable $PLR$)

$SAP$     = Source access point

$TAP$     = Target access point

$N_q$     = Qualified neighbor

## 4.1 Algorithms

| Algorithm 3: Handoff Process |
|---|

1. *CV* communicates with *SAP*
2. **if** *CV.RSSI* decreases for *T1(sec)* and $PLR_{cv} > PLR_{TH}$ **then**
3.     **for** all $N_q \in CV$ **do**
4.         *CV* disseminated *neighbor_req_message*
5.     **endfor**
6.     **if** *CV* gets *neighbor_resp_message* from $N_q$ **then**
7.         **if** $(N_q == AP_i)$ [∵ i is the id of *AP*]
8.             $TAP.CV = N_q.AP_i$
9.         **else**
10.             $CV.TAP = SAP.Vehicle_i$ [∵ i is the id of vehicle]
11.             $CV.NAV = N_q.vehicle_i$ [i.e., $NAV = N_q.vehicle_i$]
12.         **endif**
13.     *CV* waits for successful packet form *NAV*
14.         **if** (success = true) **then**
15.             **if** $PLR_{CV} > PLR_{TH}$ and $DIST(TAP_1(V)) < R$ **then**
16.                 $CV.TAP = NAV.SAP$
17.             **else**
18.                 Continue communication with *SAP*
19.                 Go to line 15
20.             **endif**
21.         **else**
22.             Go to line 1
23.         **endif**
24.     **else**
25.         Go to line 1
26.     **endif**
27. **else**
28.     Go to line 1
29. **endif**

Algorithm 3 describes various message dissemination techniques involved in the handoff process. CV gets the RSSI from SAP periodically. When the RSSI decreases continuously in a given time interval ($t$), CV calculates the PLR against SAP. If the PLR is higher than the acceptable threshold, CV starts the handoff process with the qualified neighbor vehicles through the neighbor request message. Once CV receives the neighbor reply message from the qualified neighbor, it validates the reply message to find if the neighbor is an AP or a vehicle.

If the neighbor is an AP, then CV sets the AP as the TAP and stores its IP address and MAC address to connect faster after leaving the SAP. If the source is a vehicle (NAV), then CV sets the SAP of the qualified neighbor vehicle as the TAP. CV communicates with current SAP and calculates RSSI and PLR as mentioned before. When the RSSI decreases and PLR increases than the calculated threshold, CV initiates the handoff process to transfer its connection to TAP. Thus, the actual handoff time is faster as CV receives the required information about TAP to perform handoff in a timely manner.

**Notations used in Algorithm 4:**

*SAP* = Source Access Point

$NV_i = i^{th}$ neighbor vehicle (i.e., vehicle in the communication range of ($v$))

*CV* = current vehicle

---

Algorithm 4: Neighbor assisted process

---

1. $NV_i$ receives *neighbor_req* message from *CV*
2. **if** $NV_i.SAP \neq CV.SAP$ and $NV_i$ moves in the direction of *CV* **then**
3.     $NV_i.start\ timer$()
4.     **if** ($NV_i.timeout$()) **then**
5.         $NV_i.send(Neighbor\_rep\_message)$ to *CV*
6.         $NV_i.authenticate(CV)$
7.         $NV_i.association(CV)$
8.     **else**
9.         **if** $NV_i$ receives *Neighnor_rep_message*
10.           $NV_i.stop\_timer$()
11.           $NV_i.drop\_message$
12.         **else**
13.           Go to step 4
14.         **endif**
15.     **endif**

16. **else**
17.     $NV_i.drop\_message$
18. **endif**

---

Algorithm 4 describes the neighbor-assisted process. When the qualified neighbor vehicle receives the neighbor request message from the CV, it checks the moving direction of CV (does the CV move in the same direction as the neighbor vehicle?) and current SAP of CV (does the current SAP of the CV and neighbor vehicle are same?). Once these conditions are satisfied, the neighbor vehicle starts the timer to its PLR. When the timer times out, the neighbor vehicle sends the neighbor response message to the CV, briefly explained in Sec. 4.2.

*4.2 Performance Evaluation*

Performance analysis of CVFH is calculated and compared with IEEE 802.11 protocol. The analytical model developed to compare the performance CVFH and IEEE 802.11 is discussed below:

*4.2.1 IEEE 802.11*

In IEEE 802.11, the handoff latency and successful handoff probability can be calculated as [8]:

$$T_{802-11} = \sum_{i=1}^{N_{802-11}} T_{V-I}^i + T_{auth} + T_{asso} \qquad (3)$$

$$P_{802-11} = (1 - P_{V-I}^e)^{N_{802-11}} \qquad (4)$$

Where, $T_{802-11}$ = handoff latency, $T_{V-I}^i$ = transmission time of the i$^{th}$ packet between NAV and TAP, $T_{auth}$ = authentication time, $T_{asso}$ = association time, $P_{V-I}^e$ = average packet loss rate in the communication between a vehicle and TAP, and $N_{802-11}$ = number of packets exchanged between a vehicle and TAP.

*4.2.2 CVFH*

Poisson distribution is defined as the event occurring in a fixed interval of time. In general, the Poisson distribution is given by:

$$p(k\ events\ in\ interval) = \frac{\lambda^k}{k!}e^{-\lambda} \qquad (5)$$

Where, $\lambda$ = arrival rate, $k$ = number of events, and e = Euler's constant (2.71828).

Assume that in CVFH, vehicles travelling on a road follows a Poisson distribution. The Poisson distribution of CVFH is given by:

$$p(x = k) = \frac{\lambda^k}{k!}e^{-\lambda} \qquad (6)$$

Where $\lambda$ = average density of vehicles on the road, $x$ = number of vehicles, and e = Euler's constant.

The performance evaluation of CVFH is based on the following three cases: 1) NAV and CV travelling in the same direction, 2) NAV and CV travelling in the opposite direction, and 3) Neighboring node is an TAP.

*4.2.2.1 Case 1: NAV and CV travelling in the same direction*

Assume the moving speeds of CV and NAV are represented as $V_{CV}$ and $V_{NAV}$ respectively with $V_{NAV}$ moving ahead of $V_{CV}$ in the same direction. The speed difference ($\Delta V$) between the two vehicles $V_{CV}$ and $V_{NAV}$ is given by [8]:

$$\Delta V = V_{CV} - V_{NAV} \tag{7}$$

Where, $V_{CV}$ = moving speed of CV and $V_{NAV}$ = moving speed of NAV.

From Eqn. (7), we can infer the two possible scenarios involved: 1) $\Delta V < 0$, represents the travelling speed of CV is less than the travelling speed of NAV, and 2) $\Delta V > 0$, represents the traveling the speed of CV is high compared to the travelling speed of NAV.

*Scenario 1: $\Delta V < 0$*

When $\Delta V < 0$, the distance between CV and NAV increases as the time passes by. To exchange $N^{th}_{v-v}$ packet, the distance between CV and NAV should be smaller than $R - \Delta V . T_{wl}$.

$$T_{wl} = \sum_{i=1}^{N_{V-V}} T^i_{V-V} + \sum_{i=1}^{N_{V-I}} T^i_{V-I} \tag{8}$$

Where, $R$ = communication range of a vehicle or an AP, $R - \Delta V . T_{wl}$ = maximum distance such that the vehicles CV and NAV are in the communication range of each other, $N_{V-V}$ = number of packets exchanged between CV and NAV, $N_{V-I}$ = number of packets exchanged between NAV and TAP, and $T_{wl}$ = time elapsed in wireless transmissions.

To conduct a successful handoff process, first we need to calculate the probability that at least two vehicles are in the communication range ($P'_{V-V}$). It ensures the successful transmission of packets between CV and NAV. The probability that at least two vehicles in the communication range ($P'_{V-V}$) is given by:

$$\begin{aligned} P'_{V-V} &= P(x > 1 | (R - \Delta V . T_{wl})\lambda) \\ &= 1 - P(x = 0 | (R - \Delta V . T_{wl})\lambda) - P(x = 1 | (R - \Delta V . T_{wl})\lambda) \\ &= 1 - R\lambda e^{-(R-\Delta V . T_{wl})\lambda} - e^{-(R-\Delta V . T_{wl})\lambda} \end{aligned} \tag{9}$$

Successful transmission probability of $N_{V-V}$ packets between vehicles CV and NAV ($P^S_{V-V}$) can be calculated as:

$$P^S_{V-V} = (1 - P^e_{V-V})^{N_{V-V}} \tag{10}$$

Where, $P^e_{V-V}$ = average packet loss rate between two vehicles (CV, NAV).

Successful transmission probability of $N_{V-I}$ packets between NAV and TAP ($P^S_{V-I}$) can be calculated as:

$$P^S_{V-I} = (1 - P^e_{V-I})^{N_{V-I}} \tag{11}$$

Where, $P^e_{V-I}$ = average packet loss rate between a vehicle and TAP.

The handoff probability of vehicles (CV, NAV) travelling in the same direction when $\Delta V < 0$ (i.e, distance between the CV and NAV increases as the time passes by) is based on the probability that at least two vehicles are in the communication range ($P'_{V-V}$), successful transmission probability between vehicles ($P^S_{V-V}$), and successful transmission probability between vehicles and infrastructures ($P^S_{V-I}$). It can be calculated as [8]:

$$P' = P'_{V-V} \, P^S_{V-V} P^S_{V-I} \tag{12}$$

Where, $P'$ = successful handoff probability when $\Delta V < 0$.

*Scenario 2: $\Delta V > 0$*

When $\Delta V > 0$, the distance between CV and NAV decreases as the time passes by. Probability that at least one vehicle in the communication range ensures successful transmission of packets between CV and NAV as the vehicles comes closer in a time interval ($t$) and is given by:

$$P''_{V-V} = P(x > 1 | R\lambda) \tag{13}$$
$$= 1 - P(x = 1 | R\lambda) - P(x = 0 | R\lambda)$$
$$= 1 - R\lambda e^{-R\lambda} - e^{-R\lambda}$$

Where, $P''_{V-V}$ = probability of at least one vehicle in the communication range, $\lambda$ = average density of vehicles, and $R$ = communication range of a vehicle.

The handoff probability of vehicles (CV, NAV) travelling in the same direction when $\Delta V > 0$ (i.e, distance between the CV and NAV increases as the time passes by) is based on the probability that at least one vehicle is in the communication range ($P''_{V-V}$), successful transmission probability in the communication between vehicles ($P^S_{V-V}$), and successful transmission probability in the communication between vehicles and infrastructures ($P^S_{V-I}$). It can be calculated as [8]:

$$P'' = P''_{V-V} P^S_{V-V} \, P^S_{V-I} \tag{14}$$

*4.2.2.2 Case 2: NAV and CV travelling in the opposite direction*

Case 2 represents the vehicles CV and NAV travelling in the opposite direction. The speed difference between vehicles CV and NAV can be calculated as:

$$\Delta V = V_{CV} + V_{NAV} \tag{15}$$

Where, $V_{CV}$ = moving speed of CV and $V_{NAV}$ = moving speed of NAV.

As the vehicles are moving in the opposite direction, the distance between CV and NAV will decrease as the time passes by, which is a similar scenario as CV and NAV traveling in the same direction when ΔV>0. Hence, probability that at least one vehicle in the communication range ensures successful transmission of $N_{V-V}$ packets between CV and NAV is given by:

$$P'''_{V-V} = P(x > 1|R\lambda) \tag{16}$$
$$= 1 - R\lambda e^{-R\lambda} - e^{-R\lambda}$$

Where, $P'''_{V-V}$ = probability that at least one vehicle in the communication range when the vehicles are travelling in the opposite direction.

However, as the vehicles are continuously moving in the opposite direction, CV is moving towards the TAP and the NAV is moving out of the coverage of the TAP. Hence, Probability that at least two vehicles in the communication range ensures successful transmission of $N_{V-I}$ packets between CV and NAV is given by:

$$P'''_{V-I} = P(x > 1|(R - \Delta V.T_{wl})\lambda) \tag{17}$$
$$= 1 - R\lambda e^{-(R-\Delta V.T_{wl})\lambda} - e^{-(R-\Delta V.T_{wl})\lambda}$$

Where, $P'''_{V-I}$ = probability that at least two vehicles in the communication range when the vehicles are travelling in the opposite direction.

The handoff probability of vehicles (CV, NAV) travelling in the opposite direction can be calculated as [8]:

$$P''' = P'''_{V-V} P'''_{V-I} P^S_{V-V} P^S_{V-I} \tag{18}$$

Let $P_0$ be the probability that vehicles (CV, NAV) travelling in the opposite direction and $P_l$ be the probability that vehicles (CV, NAV) is travelling in the same direction. These probabilities have two possible scenarios: 1) Δ*V*<0, represents the travelling speed of the CV is less compared to the NAV, and 2) Δ*V*>0 represents the traveling the speed the CV is high compared to the NAV.

$$P = P_0 . P'''_{V-I} + (1 - P_0)(P_l P' + (1 - P_l)P'') \tag{19}$$

*4.2.2.3 Case 3: Neighboring node is a TAP*

When the neighboring node is a TAP, CV connects to TAP directly. The successful handoff probability is given by ($P^S_{V-I}$).

The average handoff probability in CVFH can be calculated as:

$$P_{CVFH} = P_{AP} P^S_{V-I} + (1 - P_{AP})P \tag{20}$$

Where, $P_{CVFH}$ = average handoff probability of CVFH, $P_{AP}$ = probability that neighboring node is a TAP, and $P_{V-I}^s$ = successful transmission probability between vehicles and infrastructures.

The average successful handoff probability and handoff time in CVFH can be calculated as follows:

$$T_{AP} = \sum_{i=1}^{N_{V-I}} T_{V-I}^i \tag{21}$$

Where, $T_{AP}$ = average handoff time of an AP, $N_{V-I}$ = number of packets exchanged between NAV and TAP, and $T_{V-I}^i$ = transmission time of i$^{th}$ packet between NAV and TAP.

From the above equations, the average handoff time of CVFH can be calculated as [8]:

$$T_{CVFH} = P_{AP}T_{AP} + (1 - P_{AP})T_{wl} \tag{22}$$

Where, $T_{CVFH}$ = average handoff time of CVFH and $T_{wl}$ = time elapsed in wireless transmissions.

*4.3 Simulation Results*

CVFH is simulated using the ns-2 simulator and its performance is evaluated and compared with IEEE 802.11 in terms of throughput and handoff delay.

*4.3.1 Handoff Delay*

The time during which the wireless connection between a vehicle and an RSU is lost is called a handoff delay. An increase in handoff delay is observed with an increase in vehicle speed or packet rates for both the protocols. But the handoff delay of CVFH is smaller compared to IEEE 802.11 at most all simulation times as the CV in CVFH does not require authentication and association like 802.11 protocol, represented in Fig. 9. Moreover, CV acquires and stores the information about TAP if the qualified neighbor is available in its communication range.

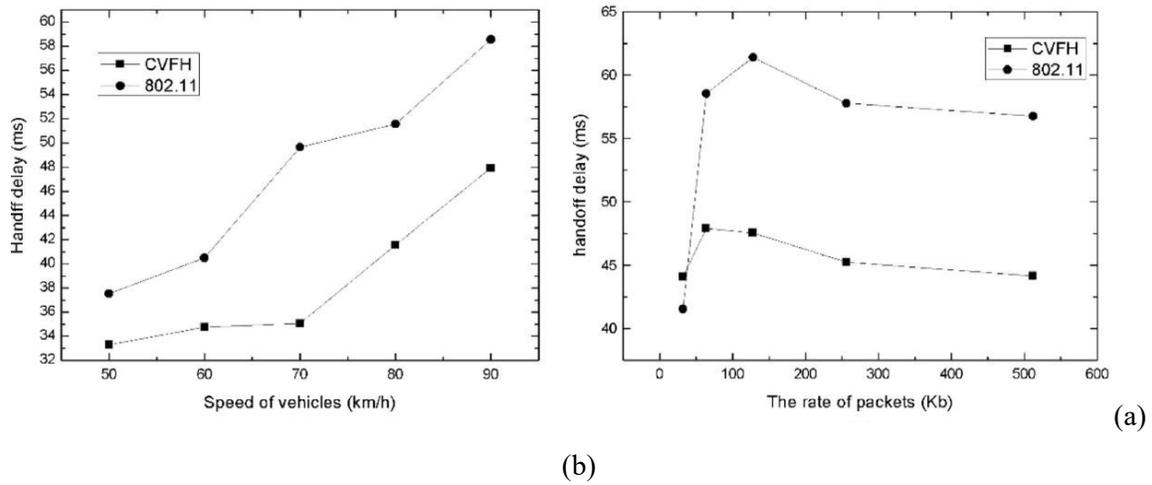

Fig. 9: Handoff delay of CVFH protocol: a) handoff delay for various speed of vehicles, b) handoff delay for various packet rates.

*4.3.2 Throughput*

A decrease in throughput is observed with an increase in speed of vehicles for both protocols because at higher vehicle speeds, the time spent for data communications with APs is less, which results in reduced throughput. It is observed that throughput of CVFH decreases faster compared to IEEE 802.11 and at vehicle speeds higher than 90 km/h a very low throughput is observed for both schemes, represented in Fig. 10. An increase in throughput is observed with an increase in packet rates. The throughput of CVFH scheme is more than IEEE 802.11 as the vehicle spends less time in handoff in CVFH.

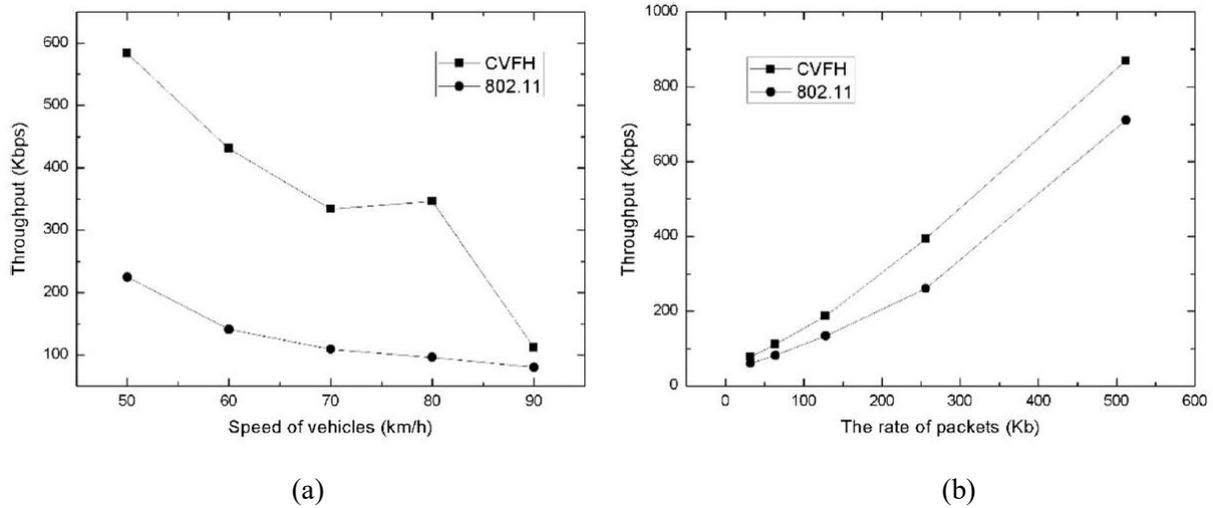

(a)          (b)

Fig. 10: Average throughput of CVFH protocol: a) average throughput for various vehicle speeds, b) average throughput for various packet rates.

## 5. FINDINGS

Two different methods of message dissemination using fog computing (i.e., FogRoute and CVFH) has their own advantages as well as pitfalls. This section presents findings of FogRoute and CVFH protocols based on the following subsections: 1) advantages, 2) limitations, 3) similarities, and 4) differences.

*5.1 Advantages*

Advantages of the FogRoute and CVFH are discussed below:

- Advantages of FogRoute is the adoption of DTN and cloud techniques to disseminate large volume of data among fog servers. The contents with the large volume such as high-quality videos, scientific datasets, etc. can be disseminated using DTN technique. Moreover, FogRoute guarantees data delivery among the fog servers by sending the required content directly or with the help of mobile devices such as vehicles, etc. connected to the fog servers. Thus, FogRoute has a high delivery ratio compared to existing DTN based data dissemination techniques.

- Advantages of CVFH is the fast handoff at low vehicle densities compared to existing handoff techniques. With the help of V2V communication and VFC, CV communicates with neighboring nodes such as vehicles or AP through request and reply messages to get the information about TAP in advance to perform the fast handoff. Thus, CVFH accomplishes less delay and high throughput to perform the handoff between SAP and TAP.

*5.2 Limitations*

Limitations of the FogRoute and CVFH are discussed below:

- Limitations of FogRoute is the order of disseminating contents to the target fog server. When a fog server determines that more than one content needs to be updated in the target fog server, fog server selects a content to be disseminated from the list of remaining contents in a random manner. Thus, the high delay associated with receiving the critical content at the target fog server. For example, disseminating the information such as road crash information, post-disaster condition plays a crucial role compared to disseminating the media content.
- Limitation of CVFH is handoff delay. It increases as the vehicles increases in the system due to a large number of packets generated being more likely to encounter another packet and resulting in frequent packet loss and retransmissions. Moreover, as the data acknowledgment message was not included in CVFH, CV will not send the acknowledgment message to NAV after receiving the information about TAP to perform handoff (i.e., reply message). Thus, when a reply message from NAV is lost (i.e., packet loss) due to the network conditions, errors in data transmission, etc. CV has to send the request message again to all the available neighbors resulting in increased handoff delay due to the dissemination of a large number of packets.

*5.3 Similarities*

The similarities between FogRoute and CVFH are discussed below:

- Both FogRoute and CVFH disseminates various messages between vehicles or between vehicles and infrastructures to perform data delivery and handoff respectively. In FogRoute, messages such as data dissemination request, data dissemination reply, etc. are exchanged between cloud provider, fog servers, and mobile devices to perform data delivery. In CVFH, neighbor request and reply messages are exchanged between nodes such as vehicles or an AP to perform the fast handoff.
- One of the main advantages of fog computing is computations are performed at the edge of networks (i.e., at the proximity of users). As FogRoute and CVFH use fog computing, the messages are disseminated rapidly between the vehicles and between vehicles and infrastructures such as RSU, etc. Thus, FogRoute and CVFH result in less delay compared to existing data dissemination and handoff techniques.

- Both FogRoute and CVFH developed an analytical model based on probability. Simulations were performed in terms of delay and delivery ratio to evaluate the performance. The results demonstrated that FogRoute and CVFH have a lower delay and higher delivery ratio in terms of data dissemination and handoff compared to existing techniques.

*5.4 Differences*

The differences between FogRoute and CVFH are discussed below:

- In most cases, FogRoute uses more than one mobile device such as vehicles to deliver high volume data such as high-quality videos, scientific datasets, etc. among fog servers. It is due to the limited storage of mobile devices. In CVFH, among all the neighbors, one neighbor (either a vehicle or an AP) replies to the request message sent by CV.
- In FogRoute, simulations were performed based on Java discrete event simulator in an urban environment with the average vehicle speed of 12.29 km/h. In CVFH, simulations were performed based on the ns-2 simulator in a highway environment with average vehicle speed of 70 km/h.
- FogRoute disseminates message between vehicles or between vehicles and infrastructures to perform data dissemination. CVFH disseminates messages between vehicles or vehicles and infrastructures to perform the handoff.

## 6. CONCLUSION AND FUTURE DIRECTIONS

In VANET, message dissemination technique plays an important role as the information between the vehicles are disseminated in the form of messages. This report discusses two existing message dissemination techniques, CVFH and FogRoute, which disseminate various messages between neighboring vehicles and fog/cloud server to perform handoff and content delivery respectively using the fog computing paradigm. We studied and analyzed the performance of CVFH and FogRoute based on the metrics delay, packet delivery ratio (also known as probability message delivery), and throughput. The results showed that the discussed protocols are efficient and perform better at most all vehicle densities. The analyses used in CVFH and FogRoute helps us to understand and design efficient message dissemination technique using fog computing.

For future work, to deliver the critical contents in a timely manner, FogRoute must include the priority module. It will allow fog servers to calculate and assign the priority for the contents needed to be updated in the target fog server, which results in content with high priority is disseminated first to the target fog server through the selected carriers (also known as the mobile device). One possible solution to reduce handoff delay of CVFH protocol is to reduce the number of packets disseminated between CV and NAV. It can be done by embedding the RFID tags on the road near the boundary region of SAP. RFID tags contain the required information about TAP to perform the handoff. Thus, when CV passes the region, RFID tags are read automatically with the help of RFID scanner and stores the information to perform handoff once CV goes out of the range of SAP.

REFERENCES

[1]   S. Ucar, S. C. Ergen, and O. Ozkasap, "Multihop-Cluster-Based IEEE 802.11p and LTE Hybrid Architecture for VANET Safety Message Dissemination," in *IEEE Transactions on Vehicular Technology*, vol. 65, no. 4, pp. 2621-2636, 2016.
      doi: 10.1109/TVT.2015.2421277

[2]   Road crash statistics. [online] Association for Safety and International Road Travel (ASIRT). Available at: https://www.asirt.org/Initiatives/Informing-Road-Users/Road-Safety-Facts/Road-Crash-Statistics [Accessed 1 Oct. 2018].

[3]   F. Cunha, L. Villas, A. Boukerche, G. Maia, A. Viana, R. A. F. Mini, and A. A. F. Loureiro, "Data communication in VANETs: Protocols, applications and challenges," in *Adhoc Networks*, vol. 44, pp. 90-103, 2016.

[4]   K. Abboud, H. A. Omar, and W. Zhuang, "Interworking of DSRC and Cellular Network Technologies for V2X Communications: A Survey," in *IEEE Transactions on Vehicular Technology*, vol. 65, no. 12, pp. 9457-9470, 2016.

[5]   X. Hou, Y. Li, M. Chen, D. Wu, D. Jin, and S. Chen, "Vehicular Fog Computing: A Viewpoint of Vehicles as the Infrastructures," in *IEEE Transactions on Vehicular Technology*, vol. 65, no. 6, pp. 3860-3873, 2016.
      doi: 10.1109/TVT.2016.2532863

[6]   C. Huang, R. Lu, and K. R. Choo, "Vehicular Fog Computing: Architecture, Use Case, and Security and Forensic Challenges," in *IEEE Communications Magazine*, vol. 55, no. 11, pp. 105-111, 2017.

[7]   L. Gao, T. H. Luan, S. Yu, W. Zhou, and B. Liu, "FogRoute: DTN-Based Data Dissemination Model in Fog Computing," in *IEEE Internet of Things Journal*, vol. 4, no. 1, pp. 225-235, 2017.
      doi: 10.1109/JIOT.2016.2645559

[8]   Bi, Yuanguo, "Neighboring vehicle-assisted fast handoff for vehicular fog communications." in *Peer-to-Peer Networking and Applications*, vol. 11, pp. 1-11, 2018.
      doi: 10.1007/s12083-017-0570-8